\documentclass[12pt,preprint]{aastex}

\slugcomment{Accepted for publication in the ApJ, August 4, 2017}

\shorttitle{Search for Binaries in PPNe}

\shortauthors{Hrivnak et al.}

\begin{document}


\title{WHERE ARE THE BINARIES?  RESULTS OF A LONG-TERM SEARCH FOR RADIAL VELOCITY BINARIES IN PROTO-PLANETARY NEBULAE}



\author{Bruce J. Hrivnak\altaffilmark{1}, Griet Van de Steene\altaffilmark{2}, Hans Van Winckel\altaffilmark{3}, Julius Sperauskas\altaffilmark{4}, David Bohlender\altaffilmark{5}, and Wenxian Lu\altaffilmark{1}}

\altaffiltext{1}{Department of Physics and Astronomy, Valparaiso University, 
Valparaiso, IN 46383; bruce.hrivnak@valpo.edu, wen.lu@valpo.edu (retired)}
\altaffiltext{2}{Royal Observatory of Belgium, Astronomy \& Astrophysics, Ringlaan 3, Brussels, Belgium; g.vandesteene@oma.be}
\altaffiltext{3}{Instituut voor Sterrenkunde, K.U. Leuven University, Celestijnenlaan 200 D, B-3001
Leuven, Belgium; Hans.VanWinckel@ster.kuleuven.be}
\altaffiltext{4}{Vilnius University Observatory, Ciurlionio 29 Vilnius 2009, Lithuania; julius.sperauskas@ff.vu.lt}
\altaffiltext{5}{National Research Council of Canada, Herzberg Astronomy and Astrophysics, 
5071 West Saanich Road, Victoria, BC V9E 2E7, Canada; David.Bohlender@nrc-cnrc.gc.ca}

\begin{abstract}

We present the results of an expanded, long-term radial velocity search (25 yrs) for evidence of binarity in a sample of seven bright proto-planetary nebulae (PPNe). 
The goal is to investigate the widely-held view that the bipolar or point-symmetric shapes of planetary nebulae (PNe) and PPNe are due to binary interactions. 
Observations from three observatories were combined from 2007$-$2015 to search for variations on the order of a few years and then combined with earlier observations from 1991$-$1995 to search for variations on the order of decades.  
All seven show velocity variations due to periodic pulsation in the range of 35$-$135 days.  
However, in only one PPN,  IRAS 22272+5435, did we find even marginal evidence found for multi-year variations that might be due to a binary companion.  This object shows marginally-significant evidence of a two-year period of low semi-amplitude which could be due to a low-mass companion, and 
it also displays some evidence of a much longer period of $>$30 years.  The absence of evidence in the other six objects for long-period radial velocity variations due to a binary companion sets significant constraints on the properties of any undetected binary companions: they must be of low mass, $\leq$0.2 M$_{\sun}$, or long period,  $>$30 years.
Thus the present observations do not provide direct support for the binary hypothesis to explain the shapes of PNe and PPNe and severely constrains the properties of any such undetected companions.

\end{abstract}


\section{INTRODUCTION}
\label{intro}

One of the outstanding problems in the study of planetary nebulae (PNe) over the past 25 years has been the determination of the mechanism or mechanisms that produce the interesting array of shapes of the circumstellar nebulae.  These range from round to elliptical to bipolar to irregular.    
This was particularly brought to attention by the intricate, often bipolar and sometimes multi-polar or point-symmetric shapes revealed by the {\it Hubble Space Telescope} ({\it HST}) \citep{sahai98,bal02}.  
Since PNe evolve from approximately spherical asymptotic giant branch (AGB) stars, this range of shapes does not meet our simplest expectation.
Initial arguments to explain these pitted rotation and magnetic fields \citep{garseg99,garseg05} versus binaries \citep{soker04,dem09} as the main driver of the shaping, and discussions of this are common at the regular series of IAU Planetary Nebulae Symposia \citep[most recently][]{liu17} and the series of Asymmetrical Planetary Nebulae conferences.

Photometric monitoring of the central stars of PNe revealed that a few of these displayed periodic photometric variability due to binarity \citep{bond00}.  As a means to investigate the role of binarity in the shaping of PNe,  photometric studies of variability in PNe increased.  With the identification by \citet{mis09} of 21 new binary central stars using the OGLE data base to study PNe in the galactic bulge,  the number of known binaries was approximately doubled.  Other studies have continued added to these \citep{jones15,hil17} and approximately 50 close binary central stars have been identified.\footnote{http://drdjones.net/bCSPN}  Most of these have periods of less than one day.
It is estimated that $\sim$20 $\%$ of PNe have close binary central stars and this could be a lower limit, as studies with the {\it Kepler} mission found several PNe with very low amplitude variability, lower than could be measured from the ground \citep{dem15}.  At least some of these are binaries.  The {\it Kepler II} mission is observing many more PNe, albeit with precision that is not as good as the earlier {\it Kepler} mission \citep{jac16}.  
Based on these short periods, it is assumed that these objects have passed through a common envelope phase.  During this time, the companion to the then AGB star was engulfed by its expanding envelope and spiraled inward, leading to the ejection of the envelope but avoiding a merger of the two stars \citep[e.g.,][]{nie12}. 
Recently \citet{hil16} presented strong 
evidence of the role of binaries in the shaping of PNe by showing the tight correlation between the angle of the orbital axis of the binary and the independently determined angle of the axis of the symmetric nebula.  Although the sample size is small, only eight objects, it consists of all of the objects which have well-determined values for the two angles and it covers a broad range of inclinations.
Central stars of PNe with wider spectroscopic orbits have recently been found \citep[e.g.,][]{vanwin14,jones17b,mis17}, but not yet in significant numbers.  There are also observational biases against discovering these, as long-term radial velocity monitoring programs are scarce and are restricted to bright central stars. 
 The role of binaries in shaping and even in forming PNe has recently been reviewed by \citet{jones17a}.

The identification of proto-planetary or pre-planetary nebulae (PPNe) in the late-1980s made it possible to investigate the precursors to PNe.  These are 
post-AGB objects in transition between the asymptotic giant branch (AGB) and the PN phase in the evolution of intermediate-mass (1.5$-$8 M$_{\sun}$) stars.  
As such, they are composed of a central star evolving toward higher temperatures that is surrounded by a detached circumstellar envelope (CSE) composed of mass lost during the AGB phase.  
The expanding gaseous component of the nebula was observed in molecular lines of CO and HCN \citep{omont93} or maser lines of OH \citep{lik89}.  Dust condensed in the outflow and gave rise to a large infrared excess that help to first identify objects as PPNe candidates \citep{hri89}.   
The dust peaked at $\sim$25 $\mu$m, at dust temperatures of 150$-$300 K.
{\it HST} observations showed that they possess faint, angularly-small nebulae (typically $<$ 4$\arcsec$), which are viewed in reflected visible light \citep{ueta00,sahai07,siod08}.
When the temperature of the central star becomes hot enough to photo-ionize this envelope, the object will have entered the PN phase.  Thus they have central stars in the range of 4,000$-$30,000 K.
The observed properties of PPNe agree well with what one would expect for the progenitors of most PNe and are expected to evolve into PNe.

Photometric studies of PPNe, particularly by Arkhipova and collaborators \citep{ark10,ark11} and Hrivnak and collaborators \citep{hri10,hri15a,hri15b} revealed light variations with periods (or quasi-periods) in the range of 35$-$160 days.  Radial velocity studies supported the idea that these variations were due to pulsations in the stars \citep{bur80,bar00,zacs09,hri13}. 

Observations of another class of stars with a infrared excess suggested that they too are post-AGB objects, with typical spectral types of F$-$G and in many cases unusual abundance patterns.  
However, in this class of stars the excess starts near the dust sublimation temperature and peaks in the near-infrared below 10 $\mu$m. 
Radial velocity studies indicated that these are binaries, with P$\approx$120$-$1500 days \citep{vanwin09}.
The dust is found to reside in a circumbinary disk \citep[e.g.,][]{vanwin03,hillen17,buj15}, and these have been referred to as post-AGB ``disc'' objects.  
They are distinguished from the PPNe, which have cooler dust not confined to a disk and are labeled ``shell'' sources. 
It seems unlikely that these objects evolve into PNe, but rather evolve into white dwarfs without passing through a PN stage.

\citet{hri11} carried out a study of seven bright PPNe to look for long-term radial velocity variations due to  binary companions.  The observations were carried out from 1991$-$1995 and then again from 2007$-$2010.  
Pulsational variations were found in each but only in one object was there a suggestion of long-term variation.
In light of the importance of verifying the presence or at least constraining the properties of any potential  binary companions to these PPNe, we have extended the study of these same seven objects over five additional years.
In this extended study, we have also increased the data density by including observations from two additional observatories, increased the precision in the observations, and more carefully considered the potential complications of comparing observations made at different times with different instruments.

\section{PROGRAM OBJECTS}
\label{object]}

The program objects are the seven brightest ({\it V}$<$10.5 mag) PPNe observable from mid-latitudes in the northern hemisphere.  Thus they can be monitored with 1$-$2 m class telescopes with high spectral resolution.
They all possess the various properties of PPNe and have been the subjects of many studies of their stellar and circumstellar properties.  
They all have spectral types of F$-$G supergiants, with double-peaked spectral energy distributions (SEDs): one peak due to the reddened photosphere and the other, in the mid-infrared, due to the re-radiation from the dust envelope \citep[e.g.,][]{hri89}. 
They are shell sources and show no evidence for the presence of a disk.
Some are oxygen-rich and some are carbon-rich, as seen in their optical spectra, infrared dust emission features, and molecular radio lines.  
Some basic properties of the program objects are listed in Table~\ref{object_list}. 

\placetable{object_list}

For each of these objects, we have obtained contemporaneous light and radial velocity curves from 2007$-$2015, and we have also obtained radial velocity curves from 1991$-$1995. 
Detailed studies of the pulsation of these have or will be published separately \citep{hri13,hri17}.

\section{RADIAL VELOCITY OBSERVATIONS AND REDUCTIONS}
\label{velocities}

We began radial velocity studies of these seven bright PPNs in 1991.  Observations were initially carried out at the Dominion Astrophysical Observatory (DAO) using the Radial Velocity Spectrometer \citep[RVS;][]{fle82} at the Coud\'{e} focus of the 1.2-m telescope.  The RVS uses a physical mask based on the spectrum of the F star Procyon and covers the spectral range 4000$-$4600 \AA.  
The incremental motion of the mask across the spectrum resulted in a cross correlation profile of the transmitted light.
The velocities were determined by fitting a parabola to the upper half of these profiles.  We will refer to these as the DAO-RVS observations. They are from 1991$-$1995, with most of the data from 1991$-$1993.\

We re-initiated the radial velocity monitoring program at the DAO in 2007, but with a CCD (DAO-CCD) replacing the RVS physical mask and photocell.  These high-resolution DAO-CCD spectra covered a smaller spectral region, 4350$-$4500 \AA.  The spectra were measured for velocity by cross correlation with a set of bright IAU radial velocity standards observed with the same instrumentation.   Again, velocities were determined by fitting a parabola to the upper half of the profiles.  More details of the DAO observations and data calibration with these two systems are given in earlier papers \citep{hri11,hri13}.
The observations reported in this study extend through the end of 2015, or to early 2016 in the case of IRAS 07134+1005.

To increase the number of observations, the observing program was expanded to include two additional telescopes.
Observations for all seven stars in this program began in 2009 on the Flemish 1.2-m Mercator Telescope on La Palma, equipped with the HERMES spectrograph.  HERMES is a fiber-fed echelle spectrograph and covers the large wavelength range of 3900$-$9000 \AA~ with a spectral resolution of $\lambda$/$\Delta\lambda$ $\sim$85,000 \citep{rask11}.  For our measurements, we used the spectral range 4770$-$6550 \AA~ (orders 55-74), which gave the fits with the best S/N.  These data were cross-correlated with a template constructed to match an F star, and the entire profile was fit with a Gaussian function to determine the velocity.  The resulting velocities are of high precision,  $\sim$0.4 km~s$^{-1}$. 
For three of the objects, observations were also carried out with a CORAVEL-type radial velocity spectrometer  mounted on the 1.65-m telescope of the Moletai Observatory (Lithuania).  The spectrum measured covered the interval 3850$-$6400 \AA, and the physical mask contains 1650 slits and was based primarily on the solar spectrum.  Velocity measurements were determined by fitting a Gaussian function to the entire profile \citep{sper16}.  Observations used in this program were carried out from 2008 to 2014.  (Earlier observations of  IRAS 22272+5435 from 2005$-$2007, with this CORAVEL-type instrument primarily at the Steward Observatory \citep{zacs09}, were not included in this study.)    
A recent study of K$-$M dwarfs with this instrument indicated a small systematic zero-point offset of 
+0.3 km~s$^{-1}$ \citep{sper16}, but we did not apply this to our F$-$G type objects (but see below).
In Table~\ref{data_sets} are listed the radial velocity data sets used in this study.
The individual radial velocities are listed elsewhere \citep{hri13,hri17}.

\placetable{data_sets}

Although the observations with the different telescope$-$spectrograph$-$detector systems were calibrated using observations of IAU radial velocity standard stars, we found that there were small but significant systematic differences between the velocity values measured.  
These were on the order of $\pm$0.5 km~sec$^{-1}$ between CORAVEL and HERMES but were larger between DAO-CCD and HERMES; on the order of  $\pm$1.0 km~sec$^{-1}$ and reached to about $+$2.0 km~sec$^{-1}$ in one case. 

We have considered in turn several possible causes for these systematic differences.
(1) These stars have complex spectra, with some containing molecules (such as C$_2$), peculiar abundances (such as enhanced s-process elements), and fast outflows.  These can effect the measured velocities when compared with more normal standard stars, depending upon whether spectral regions containing these effects are included or not.
(2) Observations made in different regions of the spectrum with the different systems would then include different regions of the complex spectra.
(3) These stars are pulsating, which produces variable asymmetries in the line profiles due to shocks and outflows \citep{leb96,zacs09,zacs16}.  Different methods of fitting the asymmetric cross-correlations function 
profiles can result in systematic differences. 
We investigated these effects empirically by analyzing the HERMES spectra over different intervals of wavelength similar to those of the other systems and indeed found systematic differences.
Thus, given these physical complexities in the stellar spectra, differences in the spectroscopic systems, and differences in the fitting of the cross-correlations profiles, it is reasonable to expect such differences at some level.

For this study, we have added empirically-determined offsets to the DAO-CCD and CORAVEL velocities to bring them to agreement with the higher-precision HERMES velocities.  The determination of these offsets is discussed in more detail in the Appendix.  These radial velocity offset values are also listed in Table~\ref{data_sets}.
Given these observed systematic differences in the radial velocity sets, we have chosen not to  include in our analysis any of the various miscellaneous velocities objects found in the literature for these seven objects.   
We do, however, refer in the Discussion to the constraints that these other observations bring to the objects' velocities during the gap in time (1996$-$2006) between our two observing sessions.

\section{INVESTIGATION OF LONG-TERM VELOCITY VARIATIONS}
\label{var}

In this study, we used our expanded data set to look for longer-term velocities variations in the following two ways.
Firstly, we combined the DAO-CCD, HERMES, and CORAVEL 2007$-$2015 data and searched for longer-term velocity variations beyond the pulsations periods over this nine-year interval.  We did this using the observational data, and where consistent pulsation periods are found, we also searched the residuals of the pulsational curves. 
Secondly, we compared the average values of the 1991$-$1995 data with the average values of the 2007$-$2015 data, looking for significant differences that might indicate variations on a time scale of decades.

The experience that we have gotten by comparing observational results from the three different telescopes-detector systems from 2007$-$2015 has also given us new insight into complications that can arise when combining the results of such different observational data sets for stars with complicated spectra.  
Hence, with these complications in mind, we did not combine the 1991$-$1995 data directly with the 2007$-$2015 data to searched for longer-term variations, as we had done in our earlier study \citep{hri11}.

\begin{figure}\figurenum{1a}\epsscale{1.0}
\rotatebox{90}{\plotone{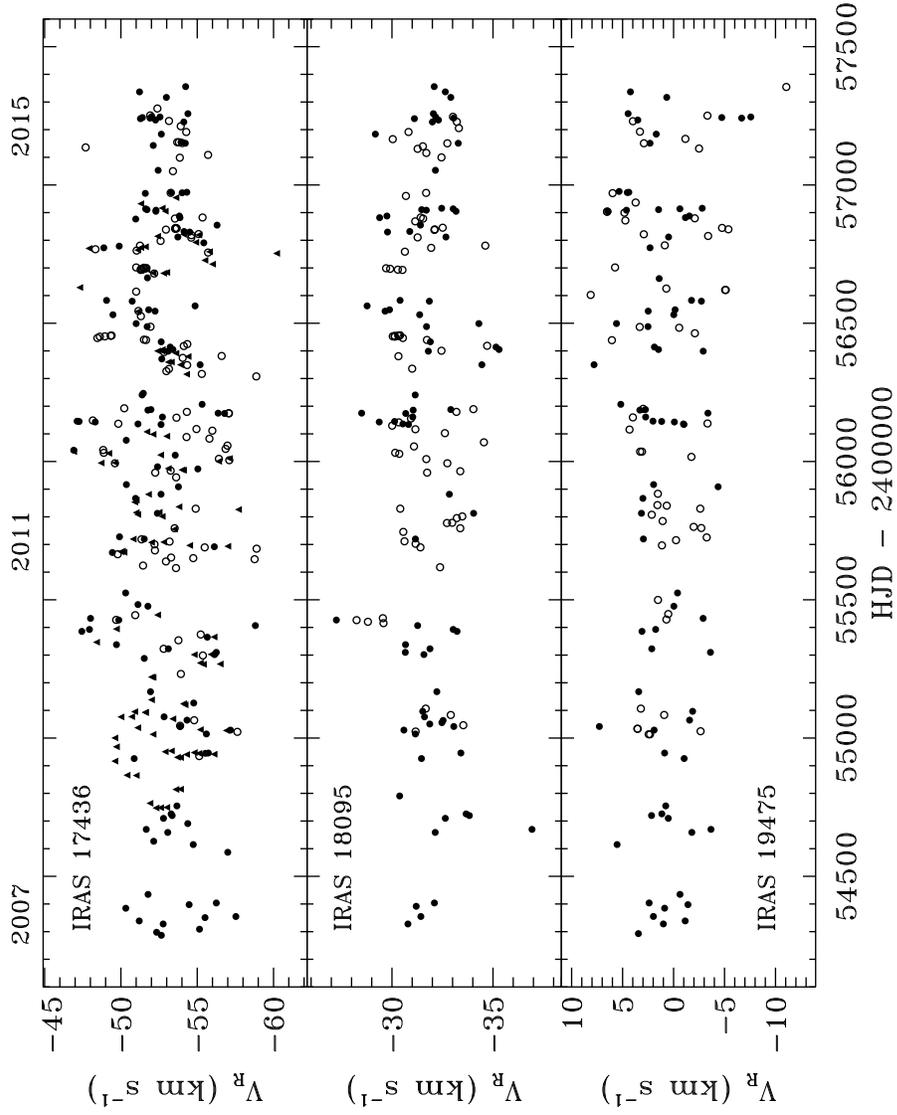}}
\caption{Radial velocity observations from 2007$-$2015.  Symbols: filled circles (DAO-CCD), open circles (HERMES), filled triangles (CORAVEL).
\label{Fig1a}}
\epsscale{1.0}
\end{figure}

\begin{figure}\figurenum{1b}\epsscale{1.0}
\rotatebox{90}{\plotone{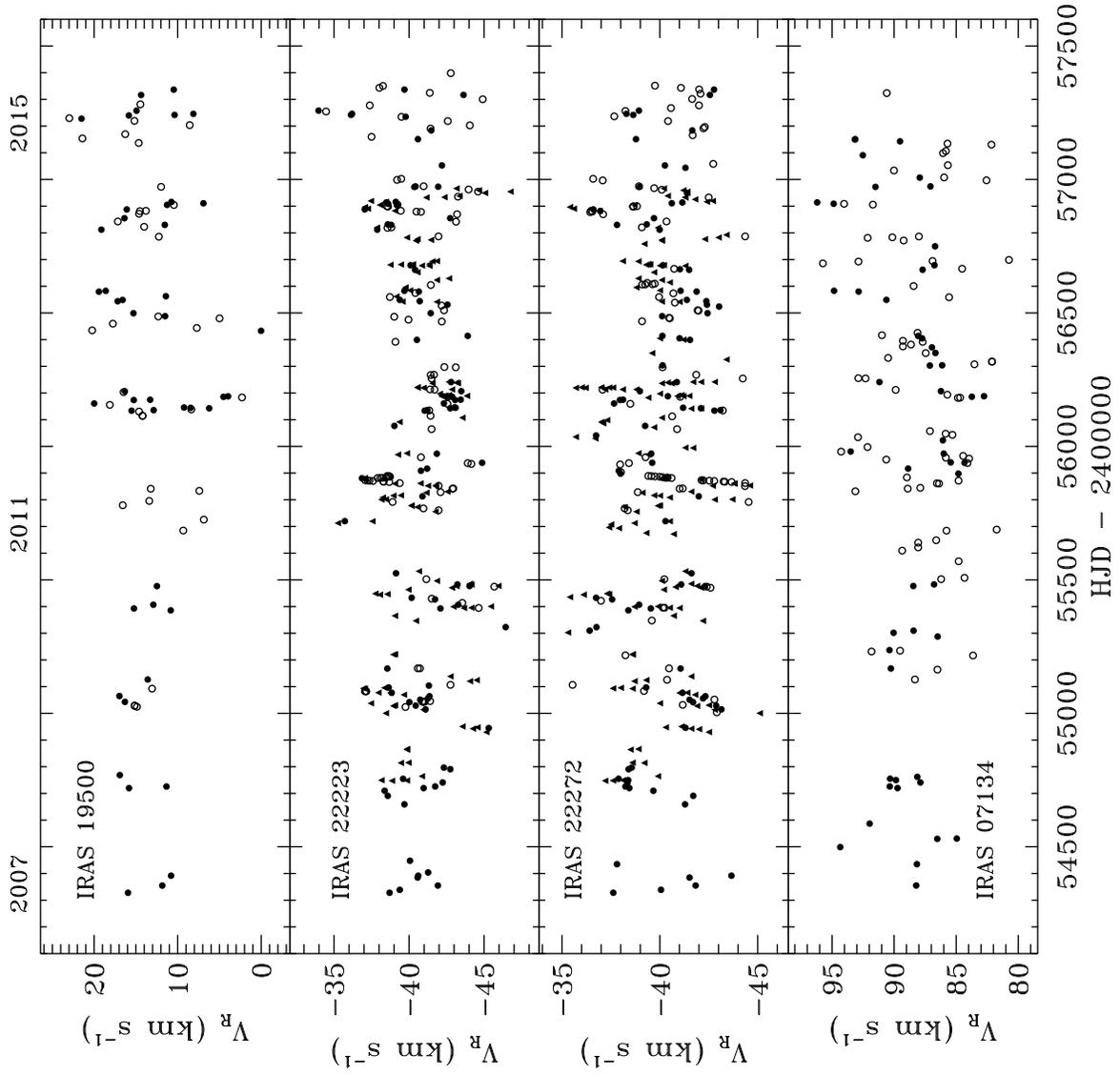}}
\caption{Radial velocity observations from 2007$-$2015.  Symbols as in Fig 1a.
\label{Fig1b}}
\epsscale{1.0}
\end{figure}

In Figure~\ref{Fig1a} and Figure~\ref{Fig1b} are shown the combined radial velocity curves for the seven objects from 2007$-$2015.  The 1991$-$1995 observations were displayed earlier \citep{hri11}.
We will discuss the objects individually in the next two subsections.
They show dominant pulsational variations in the range of 35 to 135 days, which complicate the search for binarity.
Period analyses were carried out using PERIOD04 \citep{lenz05}, which uses a Fourier technique to determine the frequencies in the data.
The recommendations of a significance criterion of a signal-to-noise ratio (S/N) greater than or equal to 4.0 in the frequency spectrum \citep{bre93} was applied. 

\placefigure{Fig1a}
\placefigure{Fig1b}

\subsection{LONG-TERM, MULTI-YEAR VARIATIONS}
\label{multi-year}

\subsubsection{IRAS 17436+5003}
\label{17436}

Radial velocity observations of this object were made from all three sites, and the combined data showed a total velocity range of 13 km~sec$^{-1}$.  
The period analysis of the three data sets from 2007$-$2015 combined led to a period of 39.4 days, but it was not formally significant.
We did, however, find that consistent periods could be found in the combined velocity curves when analyzing the data in shorter time intervals of one or two years.  These yielded periods in the range of 42$-$52 days.  
An analysis of the DAO-RVS observations yielded a period of 43 days for 1991$-$1992.
These are discussed in detail in the pulsation study by \citet{hri17}.
This object is bright, {\it V} = 7 mag, and variable, and consequently it has a long history of photometric studies, initially by Fernie 
\citep[a series of papers terminating with][]{fern95} and more recently by us \citep{hri15a}.   
Our recent photometric study revealed several periods of 42$-$50 days in the combined data.
The analyses of the velocity and light curve data document the complexities of these changing and multi-periodic pulsation curves.

We next examined the combined 2007$-$2015 radial velocity data set for periodicity longer than that of the pulsation.   The apparently complex, changing period of the short-term pulsational variations
prevented us from first removing it from the velocity curves.
Analysis of the combined 2007$-$2015 radial velocity curve yielded no evidence for a longer period in the data.

\subsubsection{IRAS 18095$+$2704}
\label{18095}

Observations from DAO-CCD and HERMES separately revealed the same period of 103.5 days, 
and the combined radial velocity curve from 2007$-$2015 resulted in a pulsation period of 103.5 days.  
This is close to the photometric period of 102.3 days found for 2010$-$2013 \citep{hri17}.
The DAO-RVS observations from 1991$-$1995 possess a period of 109.2 days, similar to the long-term (1993$-$2012) period in the light curve \citep{hri15a}.

\begin{figure}\figurenum{2}\epsscale{0.9}
\rotatebox{0}{\plotone{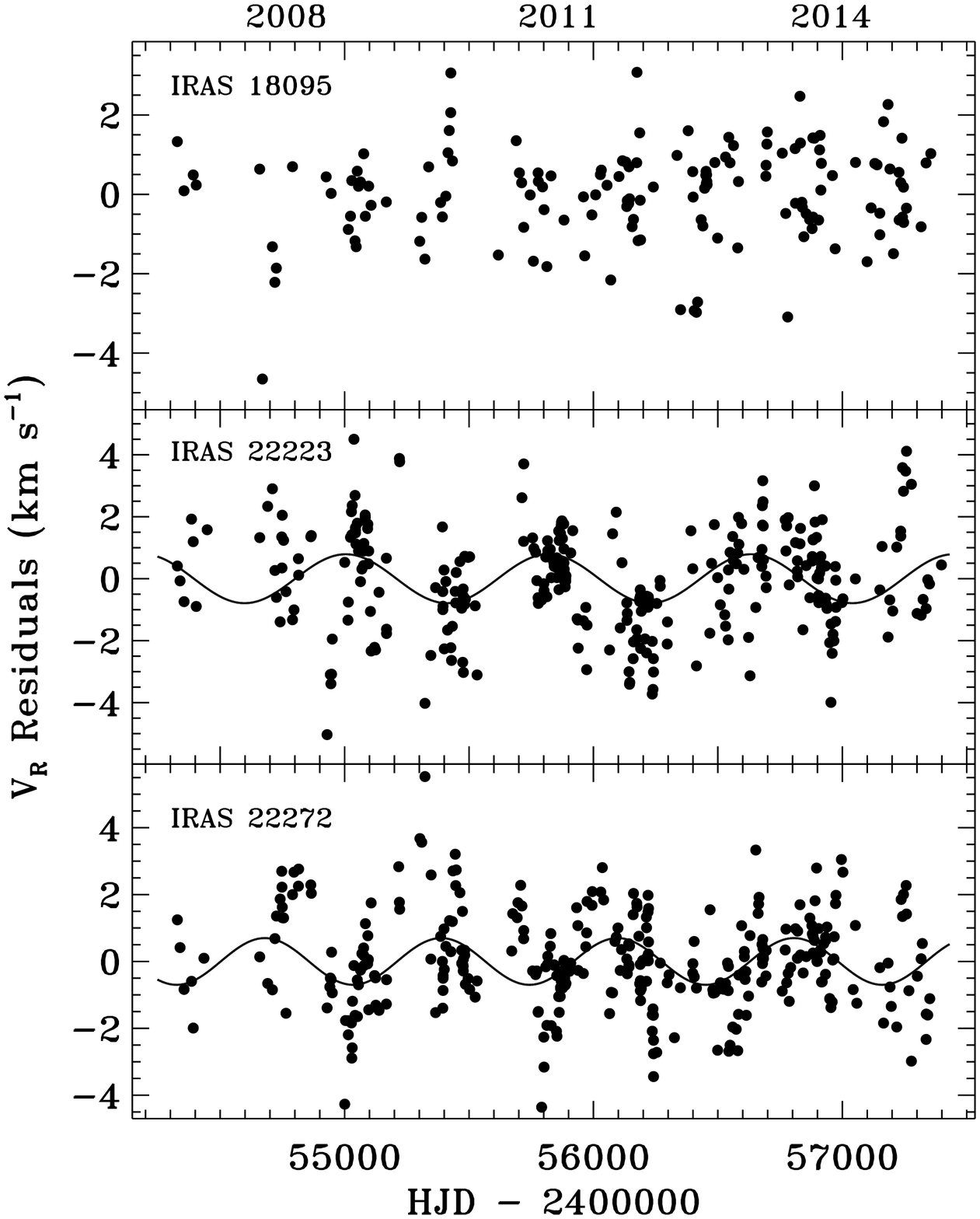}}
\caption{Residuals from the pulsation fits to the combined 2007$-$2015 data.  Shown are the sine curve fits to the residuals of IRAS 22223+43427 (P=814 days, K=0.9 km~s$^{-1}$) and IRAS 22272$+$5435 (P=710 days, K=0.7 km~s$^{-1}$).   
\label{ppne-res}}
\epsscale{1.0}
\end{figure}

The periodogram study of the combined 2007$-$2015 data did not reveal a longer period in the data over the nine seasons of observations, in either of the individual data sets or the combined data set.
Nor was one present in the residuals from the radial velocity curve when the period of 103.5 days was first removed. 
These residuals are shown in Figure~\ref{ppne-res}. 

\placefigure{ppne-res}

\subsubsection{IRAS 19475$+$3119}
\label{19475}

An analysis of the combined 2007$-$2015 radial velocity curve resulted in a period of 37.1 days.  
Periodogram analysis of the 1991$-$1995 DAO-RVS observations resulted in a significant period of 47.1 days,
although, if we restrict the analysis to the first three of  these seasons, when almost all of the observations were made, a period of 39.0 days is found. 
Periodicities in the light curve range from 35 to 43 days, similar to what is seen in the velocity curves.
We further investigated the combined data for evidence of a longer period but none was found.  
Nor was one found when we first removed the 37.1 day periodicity.  
We also found a null result when analyzing only the higher-precision HERMES data.  

\subsubsection{IRAS 19500$-$1709}
\label{19500}

This object is rather far south for observations from DAO and the observing season is relatively short from both telescopes. Consequently the number of observations from each site is more limited. 
The velocity variation for this star is larger than the others, with a range of 21 km~sec$^{-1}$.   
Separate analyses of the DAO-CCD and the HERMES velocities resulted in periods of 42.6 and 38.0 days, respectively.  Note that these values are 1-year aliases of each other.
The analysis of the combined data set yielded periods of 42.5 or 38.1 days, with the former the stronger.
The periods found from the photometric data from 2002$-$2007 are 38.3 and 42.4 days \citep{hri10}, the same periods present in the radial velocity data.
The DAO-RVS radial velocity data from 1991$-$1995 were examined and found to be fitted well with a period of 38.5 days, similar to a period found in the HERMES data and the photometry.
Thus all of the radial velocity and photometry data reveal consistent pulsation periods.
No longer-term periodicity was found in the 2007$-$2015 data set, even when we first removed the pulsation period of 42.5 days.

\subsubsection{IRAS 22223+4327}
\label{22223}

This object was observed at all three sites and the velocities yielded similar periods;
the combined data set yielded pulsation periods of 86.3 and 91.0 days.  These are similar to the periods of 86.7 and 89.7 days determined from the 2003$-$2011 light curves \citep{hri13}.
The earlier 1991$-$1995 DAO-RVS radial velocity data set has a period of 88.8 days, approximately the average value of the two periods found in the more recent data sets.  

This combined data set was examined for longer-term periodicity that might be attributed to a binary orbit.  
The periodogram analysis of the 2009$-$2015 data suggested a period of 770 days that almost met the significance criterion; it had S/N =3.9 (4.0 is the minimum considered significant).  However, when we examined the longer 2007$-$2015 data set, the resultant period, 810 $\pm$ 20 days, was less significant (S/N=3.6).  
Analyzing the velocity curves from the individual observing sites showed a period $\sim$800 days to be found in the CORAVEL data and the HERMES data, but to be formally significant only in the CORAVEL 2009$-$2014 data set.  
In Figure~\ref{ppne-res} is shown the sine curve fit to the residuals of the combined 2007$-$2015 radial velocity curve with the periods of 86.3 and 91.0 days first removed.  
The longer-term period has a velocity semi-amplitude of only 0.9 $\pm$ 0.1 km~s$^{-1}$.  
This can be compared to the full range of the combined data of 13 km~sec$^{-1}$ and the range of the residuals following the removal of the pulsations, which is 9.5 km~sec$^{-1}$.
While the fit follows the higher residuals of the 2009, 2011, 2013$-$2014 velocities and the more negative residuals of the 2010 and 2012 data, it is not a good fit to the the more positive residuals in 2008 and 2015.
Thus there is suggestion of a longer period of $\sim$800 days in the radial velocity data of IRAS 22223+4327, but it is not formally significant. 

\subsubsection{IRAS 22272+5435}
\label{22272}

This object was also observed at all three sites and similar periods were found in all three data sets.  
With the combined data set, two dominant periods of 66 and 135 days and a weaker but significant period of 162 days were determined.   
A period analysis of the 1991$-$1995 radial velocity data reveal a period of 124 days. 
The light curves have consistently shown periods of 131 and 126 days \citep{hri13}.  

Visual inspection of the 2007$-$2015 radial velocity curve shows that the average velocities are larger on some years than others, with the velocities on the even years 2008, 2010, 2012, and 2014 being more positive, both in the combined and the individual data sets.
Additional periodogram analysis of the 2007$-$2015 radial velocity curve revealed, in addition to the pulsation periods, a longer period of 710 $\pm$ 20 days that just meets our significance criterion (S/N=4.0). 
Searching for this in the individual radial velocity sets revealed a significant period of $\sim$830 days in the HERMES data, 
the suggestion of a period of $\sim$700 days in the DAO-CCD data, but no indication of such a long period in the CORAVEL data.  
Thus there is marginally-significant evidence for a period of  $\sim$2 years in the combined data from 2007$-$2015 for IRAS 22272+5435.
This is shown in Figure~\ref{ppne-res}, where we have displayed the sine curve fit of this period to the residuals of the combined velocity curve after the removal of the three shorter, more dominant periods.   
The velocity semi-amplitude is only 0.7 $\pm$ 0.1 km~s$^{-1}$. 
This is a small fraction of the full range of the combined data, 10 km~sec$^{-1}$ and of the range of the residuals following the removal of the pulsations, 8 km~sec$^{-1}$ (if we neglect one point that is $+$2 km~sec$^{-1}$ larger than the others).

The fit of this period to the seasonal variations in velocities is reasonably good, although the scatter in the residuals is still large ($\sigma$=1.3 km s$^{-1}$).  Much of this is due to the irregular amplitudes of the pulsation, which are seen in the light and velocity curves.
So for IRAS 22272+5435, there is a longer period (710 days) that is formally significant.  In Section~\ref{disc} we explore the implications of a 2-year period that may be due to a binary companion.

\subsubsection{IRAS 07134+1005}
\label{07134}

The radial velocity of this object varies over a range of 15 km~sec$^{-1}$.
When investigated separately, no significant period was found in the HERMES radial velocities, but a marginally significant period of 83.2 days was found in the DAO-CCD data.  
These data sets were then combined, although the systematic offset was rather uncertain ($+$2.0$\pm$1.0 km~sec$^{-1}$).
No significant periods were found in the combined radial velocity curve over the entire interval of 2009$-$2016, either shorter-term ones due to pulsation or longer-term ones that might indicate a binary behavior.
The 1991$-$1995 DAO-RVS data set is small, with only 21 observations and 11 of them in the 1991$-$1992 observing season.  A period of 145 days was found for the entire data set and a period of 197 days in the 1991$-$1992 season for this small sample.
Previous light and velocity curves suggested periods in the range of 35$-$45 days \citep{bar00,hri10}.

The pulsations of these low-gravity stars induce complex atmospheric differential motions.  A detailed study of the atmospheric motions of IRAS 07134+1005 was presented by  \citep{leb96} and modeled by \citep{fok01}.  These studies showed that complex atmospheric motions produce line profile asymmetries, making difficult the detection of orbital motion with a low amplitude.

\subsection{LONG-TERM, MULTI-DECADE VARIATIONS}
\label{multi-decade}

\begin{figure}\figurenum{3}\epsscale{0.75}
\rotatebox{0}{\plotone{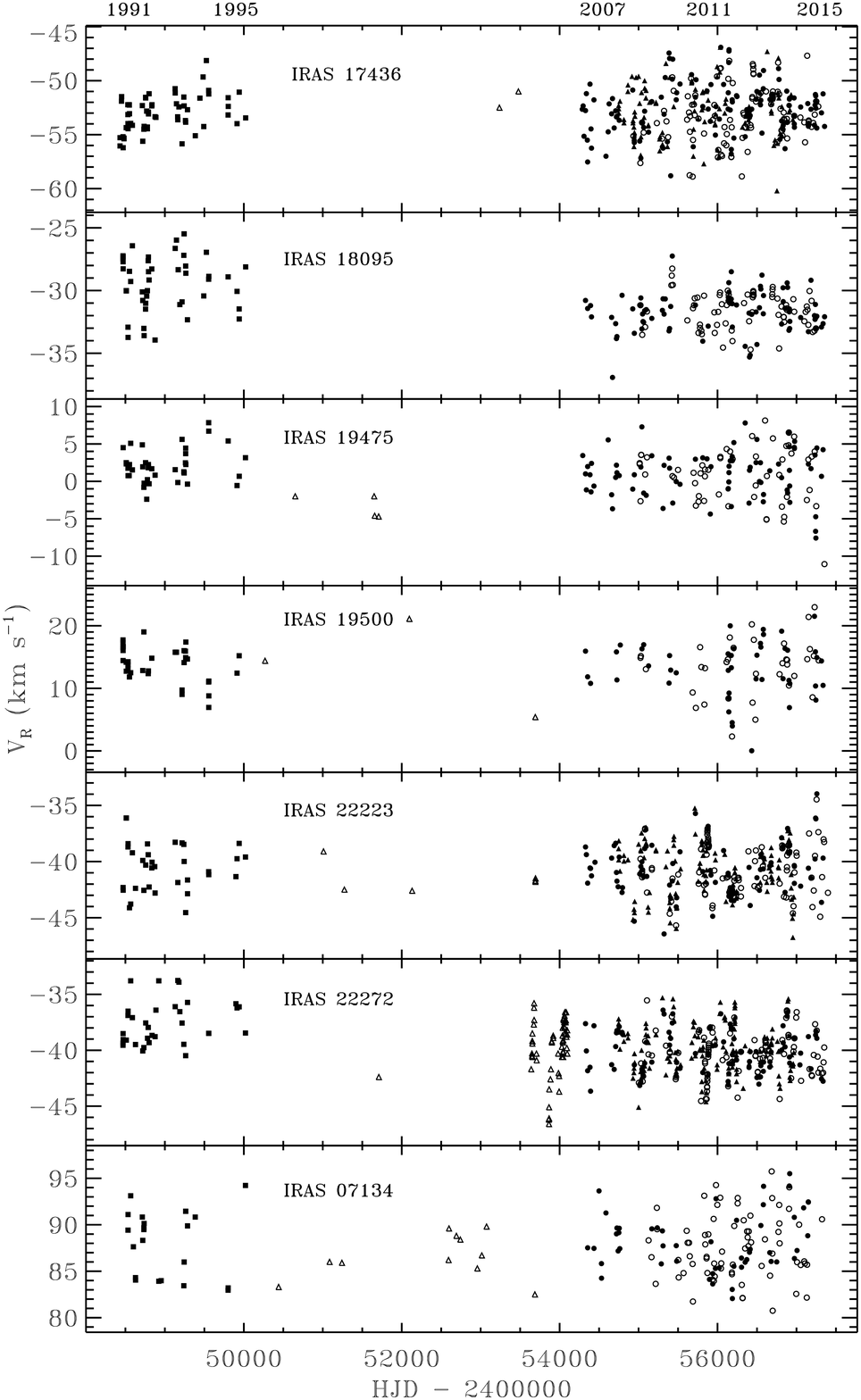}}
\caption{Long-term radial velocity study, comparing the 1991$-$1995 and the 2007$-$2015 values.  
Symbols: filled squares (DAO-RVS), filled circles (DAO-CCD), open circles (HERMES), filled triangles (CORAVEL).  Other values from the literature from 1996$-$2006 are plotted as open triangles.
\label{rvall}}
\epsscale{1.0}
\end{figure}

In Figure~\ref{rvall} is shown a composite of the long-term radial velocity data for all seven objects, 
including the 1991$-$1995 observations.
We would like to use these data to investigate longer-term, multi-decade variations in the radial velocities that might be due to a longer period binary.
We started by comparing the DAO-RVS radial velocity values from 1991$-$1995 with the HERMES values from 2009$-$2015 for the objects, to see if there were significant differences that arose during the gap in observations of 14 years.  
(The combined 2007$-$2015 values, with systematic offsets included, are similar the the HERMES values.) 
For most of the objects, we had determined periodic variations in the data sets, and so we compared the systemic velocities of the 1991$-$1995 data set with the 2009$-$2015 data set.  For the two in which we were not able to do this (IRAS 17436+5003, 07134+1005), we simply used the average values.
In Table~\ref{data_sets} (col. 7) are listed the differences, in the sense of the HERMES $-$ DAO-RVS velocities. 
They range from $-$2.8 to 0.0 km~sec$^{-1}$, and are particularly large for IRAS 22272+5435 ($-$2.8 km~sec$^{-1}$) and IRAS 18095$+$2704 ($-$2.2 km~sec$^{-1}$).  

\placefigure{rvall}

The interpretation of this comparison between the 1991$-$1995 and the 2009$-$2015 data sets is, however,  complicated by our recognition that systematic offsets can arise between different radial velocity observing systems.
This was evidenced even when comparing the HERMES, DAO-CCD, and CORAVEL data sets that were  observed contemporaneously. 
In Table~\ref{data_sets} (col. 6) are listed the systematic  velocities offsets that were determined between the HERMES and the DAO-CCD systems for each of these stars.  
As noted earlier, they range from $-$1.45 to +2.0: km~sec$^{-1}$.

Since the DAO-RVS observations were made with a different telescope-detector system at a different time than the 2007$-$2015 observations, there is no direct way to distinguish between a significant systematic offset and a long-term, multi-decade variation in the radial velocities.  
Nevertheless, we attempted to investigate this 
by comparing the DAO-RVS and DAO-CCD measurements to get a sense of whether the above differences between the HERMES and DAO-RVS measurements are larger than expected.
Suppose that we assume that the velocity offsets between the DAO-CCD and HERMES systems and the DAO-RVS and HERMES systems would be expected to be similar in value.  The basis for such an assumption might be that they both use the same DAO telescope and spectrograph, albeit with different detectors, over a relatively small wavelength range in the blue part of the spectrum.
We then compared the differences between the measured HERMES and DAO-RVS velocities (col. 7) to the previously determined HERMES and DAO-CCD offsets (col. 6). 
These differences are listed in column 8 of Table~\ref{data_sets}.
In this case, IRAS 18095$+$2704 no longer stands out, and the differences between the DAO-RVS and HERMES values are generally similar to the differences between the DAO-CCD and HERMES values (ranging from $-$0.7 to $+$1.2 km~sec$^{-1}$ except for two outliers.  
These are IRAS 22272+5435, for which the DAO-RVS difference remains large, $+$3.3 km~sec$^{-1}$, even when compared to the DAO-CCD values, and IRAS 07134+1005, for which the value is large, approximately $+$2.4 km~sec$^{-1}$, but uncertain.  

Thus we are not able to determine with certainty if the differences between the 2009$-$2015 HERMES and the 1991$-$1995  DAO-RVS measurements represent real differences in the motion of the PPNe or if they arise from systematic effects found between the different systems.
The fact that the differences are all $\le$ 0.0 km~s$^{-1}$ suggests the presence of a real, negative systematic offset.  The average difference between the Hermes and the DAO-RVS values is $-$1.0 $\pm$ 0.4 km~s$^{-1}$, changing to $-$0.7 $\pm$ 0.3 km~s$^{-1}$ if we neglect IRAS 22272+5435.
Only for IRAS 22272+5435 is the difference both well-determined and much larger than that found for any of the other well-determined empirically-measured offsets. 
Thus for IRAS 22272+5435 
there remains the suggestion of a real, multi-decadal difference in radial velocity.

\section{DISCUSSION AND CONCLUSIONS}
\label{disc}

We have carried out a long-term observational study of a sample of seven bright PPNe.  They have been examined over several timescales, and several results have been found this far.  

\vspace{-0.1in}
\begin{enumerate}
\item Pulsation periods of 35$-$135 days have been found for each of them.  

\vspace{-0.1in}
\item Evidence for multi-year periodic variations is weak, with only two of the objects showing suggestions of this (P$\approx$2 yrs) at a level just at or below the significance criteria.  And in these two cases, the amplitudes are very small ($<$1.0 km~s$^{-1}$), and the results were seen in only some of the data sets.   
Thus the PPNe clearly differ from the post-AGB binary ``disc'' sources, in whch the pulsations and binary motions are clearly distinguished and separable \citep{manick17}. 

\vspace{-0.1in}
\item Evidence for multi-decade variations is seen in one of the objects, IRAS 22272+5435.  
However, as discussed above, the interpretations of the difference between the early and later data sets is tempered by the empirical measurement 
of systematic differences (offsets) among the three recent contemporaneous data sets.
The larger, multi-decade variation observed in IRAS 22272+5435 might indeed be real, but will require continued long-term observations with the same or a similar instrument to confirm.
\end{enumerate}

What are the implications of a two-year period of small amplitude?  
In the best case for a multi-year periodic variability due to a binary companion, IRAS 22272+5435, the observed values are {\it P} = 710 $\pm$ 20 days with semi-amplitude {\it K} = 0.7 $\pm$ 0.1 km~s$^{-1}$.  The visible-band {\it HST} image of this objects shows a bipolar or multipolar nebula \citep{ueta00}.  A mid-infrared study indicates a bipolar morphology with a torus, and this has been modeled by \citet{ueta01}, who deduced an inclination of the torus of 25 $\pm$ 3\arcdeg.  
If we assume that the torus is caused by a binary companion orbiting in the same plane, then we can calculate the mass of the secondary star ({\it M$_2$}).
If we assume {\it M$_{\rm PPN}$} = 0.62 M$_{\sun}$ and a circular orbit, then the secondary has the very low mass of 0.053 $\pm$ 0.006 M$_{\sun}$.  Exploring a wide range for the mass of the PPN, one finds that for these parameters, if {\it M$_{\rm PPN}$} = 0.50 M$_{\sun}$ then {\it M$_2$} = 0.046 M$_{\sun}$, and if {\it M$_{\rm PPN}$} = 0.80 M$_{\sun}$ then {\it M$_2$} = 0.063 M$_{\sun}$.
These {\it M$_2$} values are appropriate masses for late-M spectral type dwarf stars \citep{bar96}.
If we generalize and assume that {\it K} = 1.0 km~s$^{-1}$, {\it P} = 2 years, {\it M$_{\rm PPN}$} = 0.62 M$_{\sun}$ and a circular orbit, then for various inclinations we find very low to low values for {\it M$_2$} ranging from 0.032 M$_{\sun}$ for {\it i} = 90\arcdeg, to 0.066 M$_{\sun}$ for {\it i} = 30\arcdeg, to 0.14 M$_{\sun}$ for {\it i} = 15\arcdeg.

There are also astrophysically-important implications for the radius of the orbit with a two-year period.  
Again assuming a circular orbit, the radius of the orbit is $\sim$300 R$_{\sun}$ for {\it M$_2$} within the ranges of 0.03 to 0.14 M$_{\sun}$ found above.  This would put the secondary at about or within the radius of the PPN progenitor on the tip of the AGB ($\sim$400 R$_{\sun}$), and thus could lead to a spiraling in of the secondary due to the loss of angular momentum as it orbits within the tenuous atmosphere of the AGB star \citep{nie12}.

One can also explore the evidence for multi-decadal change in the radial velocity of IRAS 22272+5435, the one object that shows a large difference in its average velocities between the recent (2007$-$2015) and the early (1991$-$1995) measurements.  
Let us assume for the moment that the change of 2.8 km~s$^{-1}$ represents a real change and is not due to a systematic effect, as has been seen at lower levels among the recent data sets.
Assuming a circular orbit, the gap of 11 years in our observations is too short for the system to have gone through a half cycle in that interval when compared to the nine year interval from 2007 to 2015 in which the observations are at a fairly constant level.  Thus the associated period must on the order of or longer than twice the time interval of 18 years between the middle points of each of the recent and early data sets.
If we assume a period of 36 years and a semi-amplitude of 1.4 km~s$^{-1}$, then for {\it M$_{\rm PPN}$} = 0.62 M$_{\sun}$ in a circular orbit with {\it i} = 25\arcdeg, {\it M$_2$} = 0.36 M$_{\sun}$.
This would imply a situation in which the secondary is of low mass and at a separation of $\sim$11 AU or $\sim$2300 R$_{\sun}$ from the PPN.

Of course, one might speculate that IRAS 22272+5435 or one of the other objects might be a binary with a very elliptical orbit, which makes an extreme excursion in velocity during the gap of 11 years (1996$-$2006) between our early and recent observations.  Such a situation was recently illustrated for the central object BD+30$\arcdeg$ 623 of the PN NGC 1514.  Observations of that object made over a two-year interval showed no evidence of binary motion while observations over an eight-year interval revealed  a nine-year period orbital period with an eccentricity of 0.46 \citep{jones17b}.
To address this possibility, we have also plotted in Figure~\ref{rvall} velocity values from the literature\footnote{IRAS 17436+5003: \citet{kip07}, \citet{tak07}; IRAS 19475+3119: \citet{kloch02}; IRAS 19500$-$1709: \citet{kloch13}; IRAS 22223+4327: \citet{kloch10}; IRAS 22272+5435: \citet{red02}, \citet{zacs09}; IRAS 07134+1005: \citet{vanwin00}, \citet{hri03}, \citet{kloch07}.} 
that fall in the gap between our early and recent velocity measurements.  These each have a precision of $\le$1 km~s$^{-1}$.
We have not used these in the analysis because we have no way to measure possible systematic effects in the velocity systems, which as we have seen might be as large as $\pm$2 km~s$^{-1}$.  Nevertheless, with that in mind they do provide some constraints on the velocities during that 11-year gap.  
These velocities are all consistant with our observations; none show a large variation that might suggest an elliptical orbit.
For IRAS 07134+1005 they fill in the gap, precluding any large excursions in velocity, and 
for IRAS 22223+4327, 19500$-$1709, 19475+3119, and 22272+5435, they partially fill in the gap, making large excursions less likely.  
They have little to no effect on the gaps in IRAS 17436+5003 and 18095+2704.
Thus while these do not rule out the possibility of an highly eccentric orbit in all cases, it is highly unlikely that elliptical orbits with periods less than 25$-$30 years exist for many, if any, of these objects.

\begin{figure}\figurenum{4}\epsscale{1.0}
\rotatebox{0}{\plotone{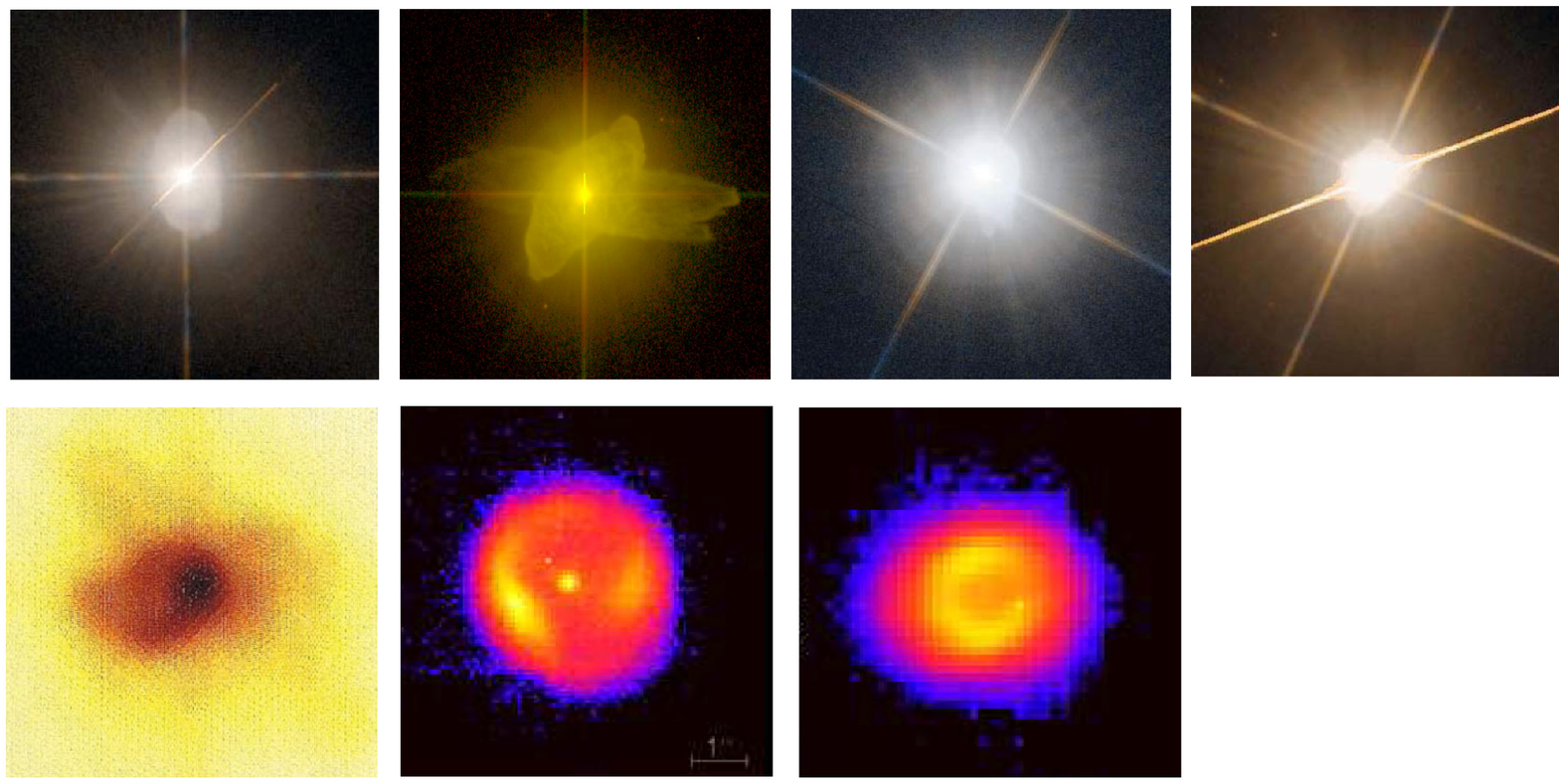}}
\caption{High-resolution visible ({\it HST}) and mid-infrared images of the seven objects in this study, taken from the literature.  
Top row (left to right): IRAS 17436+5003, 19475$+$3119, 22223+4327, 22272+5435 (Balick et al.; https://faculty.washington.edu/balick/pPNe/).
Bottom row (left to right): IRAS 18095$+$2704 \citep{ueta00}, 07134+1005, 19500$-$1709 \citep[mid-infrared;][]{lag11}.  
All of the objects appear to be bipolar or multipolar, seen at a variety of inclinations.
\label{images}}
\epsscale{1.0}
\end{figure}

 In Figure~\ref{images} are shown high resolution visible (${\it HST}$) or mid-IR images of the seven sources.  They are each clearly bipolar or multipolar with bright central stars.  
In our previous study of these seven PPNe, we discussed the possibility that our targets are biased toward low inclination as a result of our choice to observe those with bright central stars \citep{hri11}, 
since those at higher inclinations would be more likely to be obscured.
However, published modeling of their envelopes based on mid-IR observations and the appearance of the nebula in {\it HST} images indicate this not to be the case in general, and suggests that some and likely most of them are at intermediate inclinations \citep[see][for details]{hri11}.  
In this earlier study, we showed that even without knowing the inclinations of the objects, significant  limiting values of {\it P} and {\it M$_2$} could be determined for undetected companions.  These indicated that any undetected companions must have low masses ({\it M$_2$}$<$0.2 M$_{\sun}$) or be be in long-period orbits ({\it P}$>$20 years).  
These were based on assuming an upper limit for the undetected velocity semi-amplitude of 2.0 km~s$^{-1}$. 
The present study reinforces these results, implying even lower masses and/or longer periods since the upper limit of the velocity semi-amplitudes appear to be on the order of 1.0$-$1.4 km~s$^{-1}$ for a circular orbit.

\placefigure{images}

We conclude this study with the question with which we began.  Where are the binaries? 
Several alternatives suggest themselves as a result of this study, which we list below with a brief assessment.
\vspace{-0.1in}
\begin{enumerate}
\item They are present but hidden in long-period ($>$30 year) orbits $-$ perhaps. 
\vspace{-0.1in}
\item They are present but hidden due to the low mass of the secondaries ($<$0.2 M$_{\odot}$), which might even be planets $-$ perhaps. 
\vspace{-0.1in}
\item They are present but hidden inside the atmosphere of the F$-$G star $-$ unlikely due to timescale arguments for spiraling in which are on the order of the dynamic timescale of a few years, followed by the ejection of the atmosphere in a common envelope phase.
\vspace{-0.1in}
\item  They are merged with the central star $-$ but there is no evidence to support this.  They appear as normal post-AGB stars and are not rapid rotators, as documented by their high-resolutions spectral studies and the widths of their cross-correlation profiles. 
\vspace{-0.1in}
\item They are absent and these PPNe are evolving as single stars. 
\end{enumerate}
\vspace{-0.1in}

{\it The observational evidence presented in this study, based on precise radial velocity measurements of PPNe, does not provide direct support for binary companions.}
These PPNe are in clear contrast to the post-AGB ``disc'' sources studied by Van Winckel and collaborators, which all appear to be binaries and possess well-measured periods in the range $\sim$120$-$1500 days. 
It is these ``shell'' source PPNe that would be expected to evolve into PNe; however, they do not appear to be the progenitors of the short-period ($\leq$1 day) central stars found in many ($\geq$20 $\%$) PNe, which are thought to have gone through a common envelope phase.
Thus the progenitors of these short-period binary central stars of PNe remain as yet unidentified.  
Disclosing the evolutionary links between samples of known PPNe and PNe is made difficult due to the observational biases and the lack of well-constrained distances.
To prove or exclude the presence of binaries is crucial to our understanding of the way(s) that bipolar PPNe and PNe form.
If it turns out that these PPNe are not binaries, then it raises the possibility that there may be more than one way to produce bipolar PNe.
If, on the other hand, one argues that these objects do not evolve into PNe, then we are left without good candidates for the immediate progenitors of PNe.

We finally conclude this study by listing suggestions for further progress in answering this important question of the binary nature of PPNe and the necessity of a binary companion to shape the nebulae. 
One can continue to observe these seven PPNe with the same or similar instrumentation to investigate the suggested 2-year periodicity and the decadal shift in velocities, especially for IRAS 22272+5435.
This might be best carried out by using only the higher precision HERMES measurements.
One can improve our analysis with a better selection of lines to probe deeper in the star where the effects of pulsation are lessened. 
One can also observe a different sample of edge-on PPNe, where one would see the full impact of the velocity along the line of sight. Then there would be no uncertainty about the inclination of the bipolar axis with respect to the plane of the sky.  Bipolars with tight, obscured waists would particularly seem to be best explained by a binary interaction.  However, in such objects the central star is then obscured by a dusty torus.
We have initiated such a study of three edge-on PPNe for which the central star is obscured in visible light but seen in the near-infrared.  
Only a few infrared spectra have been obtained thus far and their study requires 8-m class telescopes. 

\section{APPENDIX I. EMPERICALLY-DETERMINED OFFSETS BETWEEN THE DIFFERENT RADIAL VELOCITY SYSTEMS}
\label{append1}

As we found and discussed in an earlier paper \citep{hri13}, the presence of small but systematic velocity differences exist between the different telescope$-$spectrograph$-$detector systems with their associated methods of fitting of the cross-correlation profiles, and these differ for each star.
In the present study, we therefore determined empirical adjustment offsets to correct this by comparing the measured velocities for each star in the different systems.  These were determined by three methods, and we then applied the one(s) which was most secure for a particular object.  
(1) Since these stars are known to vary in velocity over a timescale of weeks or months due to pulsation, 
we looked for instances when a star was observed with two different systems on the same or adjacent nights.  
From these we determined the differences between the observed velocities on the two systems.  
(2) We searched for periodicities in the radial velocity data sets for each star in the different systems.  Where they were found, we fitted the velocities with sine curves and then compared the systemic velocities to determine systematic differences between the two systems for that star. 
(3) We compared the average velocities for a star as observed on the different systems over the same time interval, 2009$-$2015.  
This last method gave us reasonably well-determined values for most of the objects, but since the distributions of nights are not the same, the determined systematic offsets are not as secure as the other methods for these pulsating objects, especially when there were not a lot of observations.    
This is particularly the case for IRAS 07134+1005 and 19500$-$1709; for the latter of these, the problem is exacerbated by the relatively large range of velocities for this star.

We then examined the results of the offsets determined by each of these three methods, and chose the best-determined one to use for each star.
The offsets for IRAS 18095$+$2704, 19500$-$1709, and 22223+4327 were determined by method 2. 
For IRAS 17436+5003 we used method 1, for IRAS 19475$+$3119 we used a combination of methods 2 and 3, 
and for IRAS 22272+5435 
we used the averages of the similar values determined from methods 1 and 2.
For IRAS 07134+1005, for which we found very different values from methods 2 and 3, we used a combination of the two; this one was the must unreliable. 
These empirically-determined offsets were then added to the system velocities to combine the 2007$-$2015 data for each star.
 We chose to express the values of the offsets with respect to the HERMES telescope$-$spectrograph$-$detector system because this system is the most precise and its velocities are based on a large wavelength range.
 We estimate the uncertainties in these empirically-determined offsets to be $\pm$0.7$-$1.0 km~s$^{-1}$ for the poorly determined value for IRAS07134+1005, $<$$\pm$0.5 km~s$^{-1}$ for the well-determined values for IRAS 17436+5003 and 22272+5435, and $\pm$0.5 km~s$^{-1}$ for the other four (IRAS 18095$+$2704, 19475$+$3119, 19500$-$1709, 22223+4327).

\acknowledgments  
BJH acknowledges ongoing support from the National Science Foundation
(most recently AST 1009974, 1413660).  
JS acknowledge support of the Research Council of Lithuania under the grant MIP-085/2012.
This research is based in part on observations obtained at the Dominion Astrophysical Observatory, NRC Herzberg, Programs in Astronomy and Astrophysics, National Research Council of Canada.
It is also based in part on observations made using the Mercator Telescope, operated on the island of La Palma by
the Flemish Community, situated at the Spanish Observatorio del Roque delos Muchachos of the Instituto de Astro\'{f}isica de Canarias. We used data from the HERMES spectrograph, supported by the Fund for Scientific Research of Flanders (FWO), the Research Council of K.U. Leuven, the Fonds National Recherches Scientific (FNRS), the Royal Observatory of Belgium, the Observatoire de Gen\`{e}ve, Switzerland, and the Thringer Landessternwarte Tautenburg, Germany.  This research has been conducted in part based on funding from the Research Council of K.U. Leuven (GOA/13/012) and was partially funded by the Belgian Science Policy Office under contract BR/143/A2/STARLAB.  We thank the referee for raising several points for consideration that added to the completeness of the discussion.
This research has made use of the SIMBAD database, operated at CDS, Strasbourg,
France, and NASA's Astrophysical Data System.

\clearpage

\tablenum{1}
\begin{deluxetable}{crcclrrrccl}
\tablecaption{Program Objects\label{object_list}}
\tabletypesize{\footnotesize} 
\tablewidth{0pt} \tablehead{
\colhead{IRAS ID}&\colhead{V\tablenotemark{a}}&\colhead{(B$-$V)\tablenotemark{a,b}}
&\colhead{(V$-$R${\rm _C}$)\tablenotemark{a,b}}&\colhead{Sp.T.}&\colhead{T$_{\rm eff}$}
&\colhead{log {\it g}}&\colhead{[Fe/H]}
&\colhead{[C/O]}&\colhead{Ref.\tablenotemark{c}}&\colhead{Other ID}\\
\colhead{}&\colhead{(mag)}&\colhead{(mag)}&\colhead{(mag)}&\colhead{}
&\colhead{(K)}&\colhead{}&\colhead{}&\colhead{}&\colhead{}&\colhead{}} \startdata
17436+5003 & 7.0 & 0.9 & 0.2 & F3 Ib & 6600 & 0.0 & $-$0.3 & $-$0.2 &1 & HD 161796, V814 Her \\
                   & &  & &  & 7100 & 0.5 & $-$0.2 & $-$0.3 &2 & \nodata \\
18095+2704 & 10.0 & 1.0 & 0.7 & F3 Ib & 6500 & 0.5 & $-$0.9 & $-$0.4 &3 &HD 335675, V887 Her \\
19475+3119 &  9.3 & 0.6 & 0.4 & F3 Ib & 7750 & 1.0 & $-$0.2 & $-$0.4 &4 & HD 331319, V2513 Cyg  \\
                   & & & & & 7200 & 0.5 & $-$0.2 & $-$0.5 &2 & \nodata \\
19500$-$1709 & 8.7 & 0.6 &0.4 & F3 I & 8000 &1.0 &$-$0.6 &$+$0.3 &5 & HD 187885, V5112 Sgr\\
22223+4327 & 9.8 & 1.0 & 0.5 & G0~Ia & 6500 & 1.0 & $-$0.3 & $+$0.4 &5 & BD+42~4388, V 448 Lac\\
22272+5435 & 8.6 & 2.0 & 1.0 & G5~Ia & 5750 & 0.5 & $-$0.8 &$+$0.5&6 &HD 235858, V354 Lac \\
07134+1005 & 8.2 &0.9 & 0.0 &F5 I&7250 &0.5 &$-$1.0 &$+$0.3 &5 &HD 56126, CY CMi \\
		& &&& &7250 &0.5 &$-$1.0 &$+$0.0 &7 &\nodata \\
\enddata
\tablenotetext{a}{Variable. }
\tablenotetext{b}{Includes circumstellar and interstellar reddening. }
\tablenotetext{c}{References for the spectroscopic analyses: (1) \citet{luck90}, (2) \citet{kloch02}, (3) \citet{sah11}, (4) \citet{ferro01}, (5) \citet{vanwin00}, (6) \citet{red02}, (7) \citet{red03}.
} 
\end{deluxetable}

\clearpage

\tablenum{2}
\begin{deluxetable}{clrrrccr}
\tablecaption{Radial Velocity Data Sets and Measured Velocity Differences\tablenotemark{a}\label{data_sets}}
\tabletypesize{\scriptsize}
\tablewidth{0pt} \tablehead{
\colhead{IRAS ID}&\colhead{Data Set}
&\colhead{Years} & \colhead{No.}&\colhead{$<$V$>$} 
&\colhead{$\Delta$V$_R$(offset)\tablenotemark{b}}&\colhead{$\Delta$V$_R$(HER$-$RVS)\tablenotemark{c}}&\colhead{$\Delta$V$_R$(DAO)\tablenotemark{d}}\\
\colhead{}&\colhead{}&\colhead{}&\colhead{Obs.}
&\colhead{(km~sec$^{-1}$)}&\colhead{(km~sec$^{-1}$)}&\colhead{(km~sec$^{-1}$)}&\colhead{(km~sec$^{-1}$)}} 
\startdata
17436+5003  & DAO-RVS & 1991$-$1995 & 59 & $-$53.06  & ... & 0.0 & $-$0.7 \\
                   & DAO-CCD & 2007$-$2015 & 121 & $-$51.99 & $-$0.7 &\nodata & \nodata \\
                   & CORAVEL & 2008$-$2014 & 105 & $-$53.27 & $+$0.5 &\nodata & \nodata \\
                   & HERMES       & 2009$-$2015 & 107 & $-$53.07 & ... &\nodata & \nodata \\
18095+2704  & DAO-RVS & 1991$-$1995 & 47 & $-$29.36 & ... & $-$2.2 & $+$0.8 \\
                   & DAO-CCD & 2007$-$2015 & 78 & $-$30.32 & $-$1.45 &\nodata & \nodata \\
                   & HERMES       & 2009$-$2015 & 75 & $-$31.58 & ...  &\nodata & \nodata \\
19475+3119  & DAO-RVS & 1991$-$1995 & 38 & 2.13 & ... & $-$1.0 & $-$0.2 \\
                   & DAO-CCD & 2007$-$2015 & 77 & 2.18 & $-$1.2 &\nodata & \nodata \\
                   & HERMES       & 2009$-$2015 & 57 & 0.94 & ... &\nodata & \nodata \\
19500$-$1709 & DAO-RVS & 1991$-$1995 & 35 & 13.96 & ... & $-$0.6 & $+$1.1  \\
                   & DAO-CCD & 2007$-$2015 & 46\tablenotemark{e} & 13.14 & $+$0.5 &\nodata & \nodata \\
                   & HERMES       & 2009$-$2015 & 27 & 13.84 & ... &\nodata & \nodata \\
22223+4327 & DAO-RVS & 1991$-$1995 & 34 & $-$40.52 & ... & $-$0.1 & $+$1.2 \\
                   & DAO-CCD & 2007$-$2015 & 86 & $-$41.66 & $+$1.1 &\nodata & \nodata \\
                  & CORAVEL & 2008$-$2014 & 116 & $-$41.27 & $+$0.15 &\nodata & \nodata \\
                   & HERMES       & 2009$-$2015 & 95 & $-$40.55 & ... &\nodata & \nodata \\
22272+5435 & DAO-RVS & 1991$-$1995 & 34 & $-$37.58 & ... & $-$2.8 & $+$3.3 \\
                   & DAO-CCD & 2007$-$2015 & 90 & $-$40.74 & $+$0.5 &\nodata & \nodata \\
                  & CORAVEL & 2008$-$2014 & 149 & $-$39.26 &$-$0.8 &\nodata &\nodata \\
                   & HERMES       & 2009$-$2015 & 96 & $-$40.34 & ... &\nodata & \nodata \\
 07134+1005 & DAO-RVS & 1991$-$1995 & 21 & 88.00 & ... & $-$0.4 & $+$2.4: \\
                   & DAO-CCD & 2007$-$2016 & 53\tablenotemark{f} & 86.15 & $+$2.0: &\nodata & \nodata \\
                   & HERMES       & 2009$-$2016 & 75 & 87.58 & ... & \nodata 
\enddata
\tablenotetext{a}{Uncertain values are indicated with colons (:).}
\tablenotetext{b}{The systematic radial velocity offsets determined between the HERMES and the DAO-CCD or CORAVEL velocities. (See text for details.)}
\tablenotetext{c}{The systematic differences between the HERMES and the DAO-RVS radial velocity measurements.  In the cases where the data could be fit by a periodic sine curve, we used the systemic velocities; otherwise we used average velocities.  (See text for details.)}
\tablenotetext{d}{Difference between the $\Delta$V$_R$(offset) for the DAO-CCD system and the $\Delta$V$_R$(HERMES$-$DAO-RVS) values.}
\tablenotetext{e}{Excluding one observation of IRAS 19500$-$1709 which possessed a very deviant velocity value from the others (2015; V$_R$=$-$11 km~s$^{-1}$). }
\tablenotetext{f}{Excluding five observations of IRAS 07134+1005 which possessed very discrepant values from the others. }
\end{deluxetable}

\clearpage

\clearpage

\clearpage

\clearpage

\clearpage

\end{document}